\documentclass[sort&compress,
final               
]{aipproc}

\layoutstyle{6x9}
\usepackage{amssymb,amsmath,bm,pifont}

\begin{document}
\hfill FTUV-12-08-23, RUB-TPII-03/2012


\title{Pion-photon transition form factor in light-cone sum rules
\footnote{Presented by the first author at the QCD@Work Conference, Lecce (Italy), 18--21 June, 2012.}
}


\author{A.~V.~Pimikov}{
  address={Departamento de F\'{\i}sica Te\'orica -IFIC,
                Universidad de Valencia-CSIC, E-46100 Burjassot
                (Valencia), Spain}
  ,altaddress={Bogoliubov Laboratory of Theoretical Physics, JINR,
                141980 Dubna, Russia} 
}

\author{A.~P.~Bakulev}{
  address={Bogoliubov Laboratory of Theoretical Physics, JINR,
                141980 Dubna, Russia}
}

\author{S.~V.~Mikhailov}{
  address={Bogoliubov Laboratory of Theoretical Physics, JINR,
                141980 Dubna, Russia}
}

\author{N.~G.~Stefanis}{
  address={Institut f\"{u}r Theoretische Physik II,
                Ruhr-Universit\"{a}t Bochum,
                D-44780 Bochum, Germany}
}

\begin{abstract}
We extract constraints on the pion distribution amplitude
from available data on the pion-photon transition form
factor in the framework of light-cone sum rules.
A pronounced discrepancy $(2.7-3)\sigma$
between the Gegenbauer expansion coefficients extracted from the
CELLO, CLEO, and Belle experimental data relative
to those from BaBar is found.
Predictions for the pion-photon transition form factor
are presented by employing a pion distribution amplitude
obtained long ago from QCD sum rules with
nonlocal condensates.
These predictions comply with the Belle data but disagree with those
of BaBar beyond 9~GeV$^2$.
\end{abstract}

\maketitle


\section{Pion-photon transition form factor in LCSRs}

The pion-photon transition $\gamma^{*} + \gamma^{*}\rightarrow\pi^0$
(see \cite{BL89} for a review) is
defined by the correlator of two electromagnetic currents
\begin{eqnarray}
&&  \int\! d^{4}z\,e^{-iq_{1}\cdot z}
  \langle
         \pi^0 (P)\mid T\{j_\mu(z) j_\nu(0)\}\mid 0
  \rangle
=
  i\epsilon_{\mu\nu\alpha\beta}
  q_{1}^{\alpha} q_{2}^{\beta}
  F^{\gamma^{*}\gamma^{*}\pi}(Q^2,q^2)\ ,
\label{eq:matrix-element}
\end{eqnarray}
where
$F^{\gamma^{*}\gamma^{*}\pi}(Q^2,q^2)$
is the pion-photon transition form factor (TFF) with
the photon momenta $q_1$ and $q_2$,  and
$Q^2\equiv-q_{1}^2 >0$, $q^2\equiv -q_2^2\geq 0$.
This transition process was measured by the
CELLO~\cite{CELLO91}, CLEO~\cite{CLEO98}, BaBar~\cite{BaBar09},
and Belle~\cite{Belle12} Collaborations
for the kinematics
$Q^2\gg m_\rho^2,\ \ q^2\ll m_\rho^2$ (the same also
for the planed measurement by BESIII~\cite{U12BESIII}).
For this kinematics, perturbative QCD factorization
is reliable only in the leading-twist approximation,
while higher twists are also important as shown in~\cite{RR96}.
For $q^2<m_\rho^2$, one has to include the interaction of the
(quasi) real photon at long distances $\sim O(1/\sqrt{q^2})$.
To this end, we apply~\cite{BMPS11,BMPS12}
\footnote{
This contribution is based on our previous works~\cite{BMPS11,BMPS12},
where the detailed description of our LCSR application and the
comparison of our results with some of other pion-photon TFF studies
\cite{GR08,Rad09,Pol09,NV10,RRBGGT10,ABOP10,BCT11,MS12}
can be found.
Here, we only briefly describe the basic ingredients of our
LCSR approach.
         }
Light Cone Sum Rules (LCSR)s \cite{BBK89,Kho99} that
effectively account for long-distance effects of the real photon
making use of quark-hadron duality in the vector channel
and applying a dispersion relation in $q^2$ to get~\cite{Kho99}
\begin{eqnarray}\label{eq:LCSR.FF}
  F_{\gamma\gamma^*\pi}(Q^2, q^2)
&=& %
  \int_{0}^{s_0}%
  \frac{\rho^\text{PT}(Q^2,s)}{m_\rho^2+q^2}
  e^{(m_\rho^2-s)/M^2}ds  %
  + \int_{s_0}^\infty%
  \frac{\rho^\text{PT}(Q^2,s)}{s+q^2}{ds}\, ,
\end{eqnarray}
where $s_0\simeq 1.5~\text{GeV}^2$ is the effective threshold
in the vector channel and
$M^2$ is the Borel parameter (${0.7-0.9})$ GeV${^2}$.
The effect related to the small nonzero momentum $q^2$ was
recently discussed in the framework
of LCSRs~\cite{SBMP12}, and also using Monte Carlo
simulations~\cite{CIKS12}.
Here, we neglect this effect and apply the real-photon limit
$q^2\to 0$ that can easily be taken in the LCSRs given by (\ref{eq:LCSR.FF}):
    $ F_{\gamma\gamma^*\pi}(Q^2, q^2)\equiv
      F_{\gamma\gamma^*\pi}(Q^2,0)$.

\begin{figure}[h!]
  \begin{minipage}{0.49\textwidth}
 {\footnotesize
   \begin{tabular}{|c|c|ccc|} \hline
   \hspace*{-1mm}$Q^2$\hspace*{-1mm}
               &\hspace*{-1mm}TFF\hspace*{-1mm}
                         & Tw-2      & Tw-4     & Tw-6
   \\\hline
   $3$~GeV$^2$ & $100$\% & $105.7$\% & $-17.0$\% & $11.3$\%$\vphantom{^|_|}$
   \\\cline{3-5}
               &         & LO$^\text{Tw-2}\vphantom{^|_|}$
                                     & NLO$^\text{Tw-2}$
                                                 & N$^2$LO$_{\beta_0}^\text{Tw-2}$
   \\\cline{3-5}
               &         & $130.7$\% & $-16.1$\% & $-8.9$\%$\vphantom{^|_|}$
   \\\hline
   \hspace*{-1mm}$Q^2$\hspace*{-1mm}
               &\hspace*{-1mm}TFF\hspace*{-1mm}
                         & Tw-2      & Tw-4     & Tw-6
   \\\hline
   $8$~GeV$^2$ & $100$\% & $100.5$\% & $-6.0$\% & $5.5$\%$\vphantom{^|_|}$
   \\\cline{3-5}
               &         & LO$^\text{Tw-2}\vphantom{^|_|}$
                                     & NLO$^\text{Tw-2}$
                                                 & N$^2$LO$_{\beta_0}^\text{Tw-2}$
   \\\cline{3-5}
               &         & $119.9$\% & $-13.7$\% & $-5.7$\%$\vphantom{^|_|}$
   \\\hline
   \hspace*{-1mm}$Q^2$\hspace*{-1mm}
               &\hspace*{-1mm}TFF\hspace*{-1mm}
                         & Tw-2      & Tw-4     & Tw-6
   \\\hline
   $25$~GeV$^2$& $100$\% & $99.5$\% & $-1.8$\% & $2.3$\%$\vphantom{^|_|}$
   \\\cline{3-5}
               &         & LO$^\text{Tw-2}\vphantom{^|_|}$
                                     & NLO$^\text{Tw-2}$
                                                 & N$^2$LO$_{\beta_0}^\text{Tw-2}$
   \\\cline{3-5}
               &         & $115.3$\% & $-11.9$\% & $-3.9$\%$\vphantom{^|_|}$
   \\\hline
  \end{tabular}
 }
 \end{minipage}~~~~~
 \begin{minipage}[c]{0.48\textwidth}
  \centerline{\includegraphics[width=0.9\textwidth]{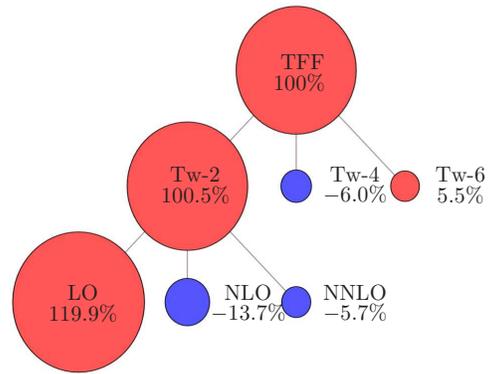}}
    \vspace{-5mm}
     \caption{Relative percentage contributions to the
     pion-photon TFF
     $F_{\gamma\gamma^*\pi}(Q^2)$ with the asymptotic
     pion DA at $Q^2=3$, $8$, and $25$~GeV$^2$
     in table form and graphically (shown only for
     $Q^2=8$~GeV$^2$).
     The blue color corresponds to negative terms, while the red one
     denotes the positive terms.
  \label{fig:LCSRterm}}
 \end{minipage}
\end{figure}

The spectral density has been calculated
in QCD up to the level of the twist-six
(Tw-6) term \cite{ABOP10}
(using similar notations for the other twists):
$$
  \rho^{\text{PT}}(Q^2,s)
=
  \frac{1}{\pi}\textbf{Im}{F_{\gamma^*\gamma^*\pi}^{\text{PT}}(Q^2,-s- i\varepsilon)}
=
  \rho_\text{Tw-2}+
  \rho_\text{Tw-4}+
  \rho_\text{Tw-6}+\ldots\, ,
$$
where the various twist contributions are given in the form of a
convolution of the hard parts with the pion distribution amplitude (DA)
of a given twist.
At the same time, for the twist-two contribution we have
the perturbative
expansion
$$
  F_{\gamma^*\gamma^*\pi}^\text{Tw-2}
\sim\left[
  T_\text{LO}  +
 a_s(\mu^2) T_\text{NLO} +
 a_s^2(\mu^2) T_{\text{NNLO}_{\beta_0}}+\ldots\right] \otimes\varphi^\text{Tw-2}_\pi(x,\mu^2)\, ,
$$
where $a_s=\alpha_s/4\pi$.
In our studies all calculated terms have been included and they
are shown in Fig.\ \ref{fig:LCSRterm} as a chain of
ovals each enclosing an area proportional to the absolute value of the corresponding
contribution.
The sum of all terms of each row gives its total contribution, as
detailed in the accompanying table in this figure.

As we see from Fig.\ \ref{fig:LCSRterm}, the twist-six
contribution~\cite{ABOP10} and the next-to-next-to-leading
order (NNLO) term~\cite{MS09} proportional to the
$\beta_0$-part have absolute values of similar size but
opposite signs in the whole $Q^2$-region.
Therefore, we consider their sum as a contribution to
the theoretical uncertainties both for fitting
the pion DA and for estimating the pion TFF
for a given pion DA model.
All other terms are included into the
central value of the TFF:
Leading order (LO), next-to-leading order (NLO) of
twist-two \cite{MS09} (corrected in~\cite{ABOP10}),
and twist-four~\cite{Kho99}.
The uncertainties of the applied LCSR approach contain
the sum of the NNLO twist-two and twist-six contributions, a 20\%
variation of the twist-four coupling
$\delta^2(\mu^2)=0.19$~GeV$^2$~\cite{BMS02},
and those uncertainties stemming from the pion DA model.

We consider the pion DA in terms of the first two coefficients\footnote{
The consideration of the pion in terms of three coefficients can be found
in \cite{BMPS11,BMPS12}.}
of the full Gegenbauer expansion $(\bar{x}\equiv 1-x)$:
\begin{eqnarray}  
  \varphi_{\pi}(x,\mu^2)
=
  6x\bar x
  \left[
  1+a_2(\mu^2)C_2^{3/2}(x-\bar x)
  +a_4(\mu^2)C_4^{3/2}(x-\bar x)
  + \ldots
  \right]
  \,. \nonumber
\end{eqnarray}
The coefficients $a_n(\mu^2)$ are evolved
according to the Efremov--Radyushkin--Brodsky--Lepage (ERBL) \cite{ER80,LB80}
evolution equation to the NLO level ~\cite{KMR86}.

\subsection{Data analysis and predictions for the TFF}

\begin{figure}[h!]
\includegraphics[width=0.65\textwidth]{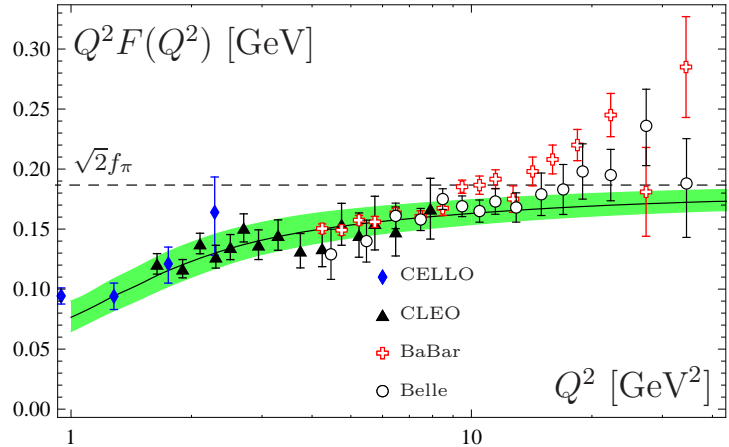}
\vspace{-5mm}
\caption{
    Theoretical predictions extracted in \cite{BMPS11,BMPS12}
    for the scaled
    $\gamma^*\gamma\pi^0$
    TFF, calculated in the LCSR approach with pion DAs
    obtained in QCD sum rules with nonlocal condensates \protect\cite{BMS01},
    in comparison with the experimental data from
    CELLO~\cite{CELLO91}, CLEO~\cite{CLEO98}, BaBar~\cite{BaBar09}, and
    Belle~\cite{Belle12}.
\label{fig:BMSvsData}
\vspace{0mm}}
\end{figure}

The described LCSR-based techniques allow the computation of
the TFF for any pion DA given in the form of a Gegenbauer expansion.
Using LCSRs with pion DAs obtained in QCD sum rules with nonlocal
condensates~\cite{BMS01}, one obtains the predictions shown by a
shaded (green) band in Fig.\ \ref{fig:BMSvsData} in
comparison with the available experimental data.

\begin{figure}[h!]
\includegraphics[width=0.75\textwidth]{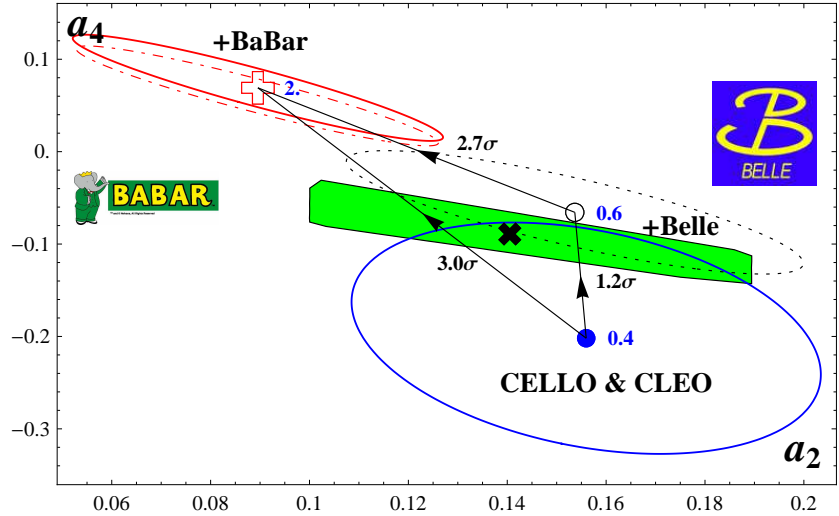}
\vspace{-5mm}
\caption{
Confidence regions for two-parameter  ($a_2,\,a_4$) fits based on
three data sets:
first - CLEO, CELLO \cite{CELLO91,CLEO98} (solid (blue) ellipse
at the right bottom);
second - CLEO, CELLO, and BaBar \cite{BaBar09} (solid (red) ellipse
at the left top);
third - CLEO, CELLO, and Belle \cite{Belle12} (dotted ellipse
at the center).
Best fit values of the $\chi$-squared goodness of fit
$\chi^2/ndf$
are shown at the apexes of a triangle.
On its sides we display the discrepancy among the
data sets in terms of a standard deviation (1$\sigma\approx 68\%$).
The area of the $a_2$, $a_4$ values allowed by QCD sum rules with
nonlocal condensates from \protect\cite{BMS01} is indicated by a
slanted (green) rectangle.
\label{fig:2D-regions}
\vspace{0mm}}
\end{figure}

On the other hand, one can fit the pion DA Gegenbauer coefficients
to the different data sets in order to obtain corresponding
experimental constraints.
This can also reveal the extent of the
discrepancy among the various data sets.
In Fig.\ \ref{fig:2D-regions}, we present the confidence
region of $a_2$ and $a_4$ given in the form of error
ellipses derived from fitting different sets of data.
Best-fit values of the $\chi$-squared goodness of fit
$\chi^2/ndf$
(ndf=number of degrees of freedom)
are shown at the centers of the ellipses, while deviations between
particular data sets are displayed on the sides of the
triangle in units of one standard deviation (1$\sigma\approx 68\%$).
From Fig.\ \ref{fig:2D-regions}, we conclude that the inclusion
of the BaBar data to those of CELLO\&CLEO leads to a
3$\sigma$ shifting of the confidence  region from the (blue) ellipse
at the bottom on the right to the (red) solid-lined ellipse at the top
on the left, accompanied by a significant increase of the
$\chi$-squared goodness of fit value from 0.4 to 2.
The additional inclusion of the Belle data plays no role in the
two-parameter analyses (dashed-dotted (red) ellipse).
If we include to the CELLO\&CLEO data only the
Belle data (ignoring the BaBar data), then the shifting
of the confidence region is moderate (1.2$\sigma$), with
a slight increase of the $\chi$-squared goodness of fit
value from 0.4 to 0.6 (dotted black ellipse).
Note that we considered the $\chi$-squared goodness of fit
for the analyzed experiments,  $\chi^2$,
as the sum of the individual $\chi$-squared goodness of fit values
associated with each experiment.
When the data strongly deviate from each other, this way of
combining data could be inefficient for extracting the confidence
region of the fitting parameters, though it is still good for
identifying and rating this deviation.

From Fig.\ \ref{fig:2D-regions}, we conclude that the BaBar data
deviate from the other data sets at the 3$\sigma$ level and cannot be
satisfied by pion DAs based on models with only two Gegenbauer
coefficients---unlike all other data sets---and hence
they require (at least) a sizable coefficient $a_6$
or still higher coefficients.

\section{Conclusions}
\label{sec:concl}
It is not a trivial matter that QCD has come to use collinear
factorization to describe basic exclusive processes, like the pion-photon
transition form factor.
This scheme has been challenged by the BaBar data because they indicate
an auxetic behavior at $Q^2 > 9$~GeV$^2$ that cannot be uniquely described
by a known QCD mechanism.
Therefore, the Belle data, which do not replicate this increase of the
scaled pion-photon transition form factor, may help restoring the confidence
to the standard QCD scheme, though a definitive answer requires more accurate
data in the high-$Q^2$ range.


\begin{theacknowledgments}
One of us (A.V.P.) is thankful to V.~Braun, and  D.~Melikhov for fruitful
discussions during the QCD@Work Conference, 2012, Lecce, Italy.
This work was supported in part by the Heisenberg--Landau Program under
Grant 2012, the Russian Foundation for Fundamental Research
(Grant No.\ 12-02-00613a), and the BRFBR--JINR Cooperation Program under
contract No.\ F10D-002.
The work of A.V.P. was supported in part
by the Ministry of Education and Science of Russian Federation, project 14.B37.21.0910,
and by HadronPhysics2, Spanish Ministerio de Economia y Competitividad and EU
FEDER under contract FPA2010-21750-C02-01, AIC10-D-000598, and
GVPrometeo2009/129.
\end{theacknowledgments}


\bibliographystyle{aipproc}

\end{document}